# Density Functional Theory-based Electric Field Gradient Database

Kamal Choudhary[1,2], Jaafar N. Ansari[3], Igor I. Mazin[3,4], Karen L. Sauer[3,4]

1. Materials Science and Engineering Division, National Institute of Standards and Technology, Gaithersburg, MD 20899, USA.
2. Theiss Research, La Jolla, CA 92037, USA.
3. Department of Physics and Astronomy, George Mason University, Fairfax, VA 22030, USA.
4. Quantum Science and Engineering Center, George Mason University, Fairfax, VA 22030, USA.

**Abstract**

The deviation of the electron density around the nuclei from spherical symmetry determines the electric field gradient (EFG), which can be measured by various types of spectroscopy. Nuclear Quadrupole Resonance (NQR) is particularly sensitive to the EFG. The EFGs, and by implication NQR frequencies, vary dramatically across materials. Consequently, searching for NQR spectral lines in previously uninvestigated materials represents a major challenge. Calculated EFGs can significantly aid at the search's inception. To facilitate this task, we have applied high-throughput density functional theory calculations to predict EFGs for 15187 materials in the JARVIS-DFT database. This database, which will include EFG as a standard entry, is continuously increasing. Given the large scope of the database, it is impractical to verify each calculation. However, we assess accuracy by singling out cases for which reliable experimental information is readily available and compare them to the calculations. We further present a statistical analysis of the results. The database and tools associated with our work are made publicly available by JARVIS-DFT (https://www.ctcms.nist.gov/~knc6/JVASP.html) and NIST-JARVIS API (http://jarvis.nist.gov/).

**Corresponding author:** Kamal Choudhary (kamal.choudhary@nist.gov)

**Background and Summary**

Nuclear resonance spectroscopies, such as Nuclear Magnetic Resonance (NMR)[1,2] and Nuclear Quadrupole Resonance (NQR)[3] are extremely valuable as sensitive probes of the local electronic structure in solids. They are considered to be the golden standard in addressing such disparate physical properties of materials such as magnetism, charge ordering, structural distortion, valence skipping, superconductivity, and many others. Shifts of the NMR/NQR spectral lines yield information of uniform susceptibilities, while their relaxation time informs about the local susceptibility at a nucleus. The very low excitation energies of nuclear resonances (even on the scale of the superconducting gap) provide a window into dynamical effects. Further, nuclear resonances have a plethora of practical applications, ranging from medical (MRI) and pharmaceutical[4-6] to detecting prohibited substances[7-9] and combatting terrorism[10-13].

The difference between NMR and NQR is that in NMR, the separation of the nuclear levels is predominantly affected through external or internal magnetic fields, while in NQR it comes from the interaction of the nuclear quadrupolar moment with the gradients of the static electric field at the nucleus (Electric Field Gradients, EFG). EFG can also be obtained from Mossbauer spectroscopy[14], however, NMR and NQR have recently become more commonly used[15] especially for crystalline solids. Each method has its advantages and disadvantages. Positions of NMR lines are determined by the nuclear magnetic moment and the field at the nucleus and are rather well known, up to small shifts induced by embedding the nucleus into a crystal, an item of principal interest in NMR. As opposed to NQR, NMR does not require a long-range order and can be applied to liquids and glasses. However, not all nuclei are NMR-active, and often the experiments require high-field magnets and cumbersome equipment. It is for this latter reason that NQR, not NMR, is used for the detection of explosives. NQR activity is present in any nucleus with spin greater than ½, a non-zero electric quadrupole moment, and located in a lower-than-cubic symmetry environment. Moreover, in some, but not all, as shown in our analysis below, the quadrupolar splitting is larger than the Zeeman splitting achievable in conventional NMR. A major disadvantage of NQR is the necessity to search over extremely wide spectral windows (tens of MHz or more) to find the resonances in the absence of a priori information on the magnitude of the quadrupolar interaction in a given material[16]. Fortunately, many modern commercial and open-access software packages for electronic structure calculations can compute the EFGs from first principles. However, by far not all experimental groups have access to such codes, and expertise to run them.

The main goal of this work is to provide the calculated values of EFGs for different compounds in one of the most comprehensive materials databases. The paper is organized as follows: first, we present, for consistency, the general theory and methodology. Next, we shall compare the computational results for cases where reliable experimental data exist, with the experiment, and discuss several individual cases where unusually large deviations have been found. In doing that, we will point out potential factors that may render EFGs particularly sensitive to computational details. Lastly, we shall discuss the general statistical distribution of the EFG parameters.

**Methods**

**General Theory**

The key parameters used to define NQR spectral lines are the quadrupole coupling constant $\nu_Q = eQV_{zz}/h$ and the asymmetry parameter

$$\eta = (V_{xx} - V_{yy})/V_{zz}, \quad (1)$$

where $e$ is electric charge, $h$ is Planck's constant, and $Q$ is the nuclear quadrupole moment; $V_{ii}$ are the principal components of the diagonalized EFG tensor, defined as the second derivative in Cartesian coordinates of the Coulomb potential at the nucleus position. By construction, the EFG, $V_{ii}$ is a traceless tensor. The coordinate system, in accordance with the convention used by experimentalists is chosen so that $|V_{zz}| \geq |V_{yy}| \geq |V_{xx}|$, which forces $0 \leq \eta \leq 1$. Note that if the point group of the site in question is cubic, then by symmetry all components are zero; if it is tetragonal or hexagonal, then $\eta = 0$, but $V_{zz} \neq 0$. Density functional theory[17] calculations can be used to predict the EFGs, however, a systematic database of such quantities for materials is still missing. Computationally, it is much easier to provide such properties for thousands of materials in a systematic way than to do so through experiments.

The Materials Genome Initiative (MGI)[18] based projects such as AFLOW[19], Materials-project[20], Khazana[17], Open Quantum Materials Database (OQMD)[21], NOMAD[22], Computational Materials Repository (CMR)[23], NIMS[24] and NIST-JARVIS[25-36] have played key roles in the generation of electronic-property related databases, and it is an obvious next step to extend them to nuclear physics-related quantities such as EFGs. There has been some systematic experimental database development in the past such as Japan Association for International Chemical Information (JAICI) Nuclear Quadrupole Resonance Spectrum (NQRS) database[37,38] which hosted NQR/NMR data with specific compound-related data for hundreds of materials but has now gone offline for public usage. However, even if an open-access comprehensive database of all known experimental data on EFG existed, it would still not solve the problem of a time-consuming experimental search for NQR lines in yet unexplored materials.

DFT appears to be a good tool, with sufficient predictive power, to point the experiment on a new compound in the right direction. Indeed, ideally, DFT, by construction, provides the exact charge density and therefore exact EFGs. In practice, of course, various approximations are used, but, as a rule of thumb, DFT is much better suited for calculating the total energy and the total charge density than for calculating, say, electron excitation spectra.

Initially, DFT was applied to EFGs in its all-electron formulation. One of the first full-potential, all-electron codes to implement this calculation was the WIEN2k package developed by Schwarz, Blaha et al. [39] implementing the full-potential linearized augmented-plane-wave (FLAPW) method. In Ref. [39] it was first applied to calculating EFGs and since then has been used for several classes of materials. It is essential for this task to have no additional approximations for the charge density and potential shape, as, for instance, in earlier muffin-tin versions of these codes. On the other hand, pseudopotential methods do not have any restrictions on the shape of the potential, being just plane wave expansions, and are very fast. The problem with such methods is that inside the atomic core they used pseudo-wavefunctions and pseudo-density, rather than the actual electronic charge. This problem was resolved with the invention by P. Blochl in 1994 of the projector augmented waves (PAW) method[40], which allows rigorous extraction of the true electronic density from PAW pseudopotential calculations. The formalism was tested by Petrilli et al.[41] and found to be consistent with all-electron calculations. Later PAW pseudopotentials were implemented in the popular software package VASP (Vienna ab-initio simulation package) and the formalism of Ref. 41 was implemented as well. This package will be used throughout this paper. A more detailed review of predicting nuclear quantities using density functional theory based methods can be found elsewhere[42,43].

In this work, we apply the PAW formalism to develop a computational database of EFGs for ~15000 materials and make the database publicly available through the NIST-JARVIS platform. We compare a few of the predicted data with experiments to estimate the uncertainty in predictions and carry out correlation and trend analysis to reveal the underpinning physics. The NIST-JARVIS (https://jarvis.nist.gov/) has several components, such as JARVIS-FF, JARVIS-DFT, JARVIS-ML, JARVIS-STM, JARVIS-Heterostructure and hosts material-properties such as lattice parameters[26], formation energies[28], 2D exfoliation energies[25], bandgaps[29], elastic constants[26], dielectric constants[29,44], infrared intensities[44], piezoelectric constants[44], thermoelectric properties[31], optoelectronic properties, solar-cell efficiencies[30,33], topological materials[28,29], and computational STM images[32]. The JARVIS-DFT (https://www.ctcms.nist.gov/~knc6/JVASP.html) currently hosts ~40000 3D, ~1000 2D materials with millions of calculated material-properties. We believe the EFG database along with other property-data will be a useful resource for material-design and can serve as precursors for artificial intelligence and data-mining methods.

**Density functional theory calculations**

The DFT calculations were carried out using the Vienna Ab-initio simulation package (VASP)[45,46]. The entire study was managed, monitored, and analyzed using the modular workflow, which we have made available on our github page (https://github.com/usnistgov/jarvis). Please note that commercial software is identified to specify procedures. Such identification does not imply endorsement by the National Institute of Standards and Technology. We use the projected augmented wave method[40,47] and OptB88vdW functional[48], which gives accurate lattice parameters for both van der Waals (vdW) and non-vdW solids[25,26]. Both the internal atomic positions and the lattice constants are allowed to relax in spin-unrestricted calculations until the maximal residual Hellmann–Feynman forces on atoms are smaller than 0.001 eV Å$^{-1}$ and energy-tolerance of $10^{-7}$ eV. We do not consider spin-orbit interactions or magnetic orderings besides ferromagnetic, because of a high computational cost. We note that nuclear spins are not explicitly considered during the DFT calculations. The list of pseudopotentials used in this work is given on the github page. The k-point mesh and plane-wave cut-off were converged for each material using the automated procedure described in Ref[36]. After optimization of the structures, we calculate the EFGs at the positions of atomic nuclei following Petrilli et al. [41]. We add 30% extra plane-wave cut-off on top of that obtained from automatic convergence script and impose a tighter electronic convergence threshold of $10^{-8}$ eV for the EFG calculations.

Starting from ~40000 3D materials the JARVIS-DFT database, we screen out materials with point group 23, $\bar{4}3m$, $\bar{m}3$, 432, m$\bar{3}$m in which atomic sites have cubic symmetry, leaving 25931 candidates. Next, we choose materials with 20 or lesser atoms in a unit cell for computational cost reasons leading to 18672 materials. Out of these candidates we have calculated EFGs for 15187 materials and EFG data for other materials would also be available soon.

**Data records**

After the calculations, the metadata is stored in the Javascript Object Notation Files (JSON) format which can be easily integrated with databases such as MongoDB. The dataset would be made publicly available through JARVIS-DFT (https://www.ctcms.nist.gov/~knc6/JVASP.html) website and NIST-JARVIS REST-API (https://jarvis.nist.gov/). We would also make a Comma Separated Values (CSV) format data and make it publicly available through Figshare repository (https://doi.org/10.6084/m9.figshare.12307700). Note in the CSV file, $\eta$ is listed as zero when $V_{zz}$ is zero, as is standard for VASP output. The CSV file contains the following entries:

Table 1 Keys for the metadata and their descriptions.

| Key | Description |
| --- | --- |
| **JARVIS-ID** | JARVIS-DFT calculation identifier |
| **Formula** | Chemical Formula |
| **Spacegroup** | Crystallographic spacegroup number |
| **Atom** | Name of the atomic site |
| **Wycoff** | Wycoff site label |
| **V$_{ZZ}$** | The largest in amplitude EFG principal component |
| **V$_{XX}$** | The smallest in amplitude EFG principal component |
| **V$_{YY}$** | The third EFG principal component |
| **Eta** | Asymmetry parameter |

**Technical validation**

As mentioned previously, the exact DFT would give the exact EFG, as opposed to, for instance, the excitation gap. In reality, however, there are several sources of inaccuracy, as discussed below.

The first limitation is that in all calculations, for the purpose of uniformity, we have used the calculated crystal structure, even for those cases where an experimental structure was available. Normally that does not have a large effect. Recall that our goal is not to predict the EFGs with maximal accuracy, but to give experimentalists a reference point for each material, in the vicinity of which they should search for spectral lines.

Notable exceptions occur when the local environment is very close to cubic, but not exactly. Consider an example of a hexagonal closed packed (hcp) material. If the ratio *c/a* is equal to its ideal value, $\sqrt{8/3} \approx 1.633$, the nearest neighbors around each site form an ideal dodecahedron and the EFGs from the six in-plane neighbors and six nearest-plane neighbors cancel out exactly. In a simple point charge model, neglecting the second and farther neighbors (whose contribution is at least 4.5 times smaller), the net contribution to $V_{zz}$ is proportional to $\left(\frac{c}{a} - \sqrt{\frac{8}{3}}\right)/\sqrt{6}$; that is to say $V_{zz}$ is linear with the deviation of $c/a$ from the ideal value, with the zero intercept very close to the latter. Indeed, we observe this linear behavior for α-Sc (Fig. 1). Note that experimentally, $c/a$ for Sc is 1.6, less than 2% away from the ideal ratio, so a 2% error in determining $c/a$ can lead to a 100% error in EFG. Other materials' sites close to a local cubic environment, where a particular caution needs to be exercised, include hcp and its derivative, including wurtzite or lonsdaleite, or structures derived from cubic, such as distorted perovskites.

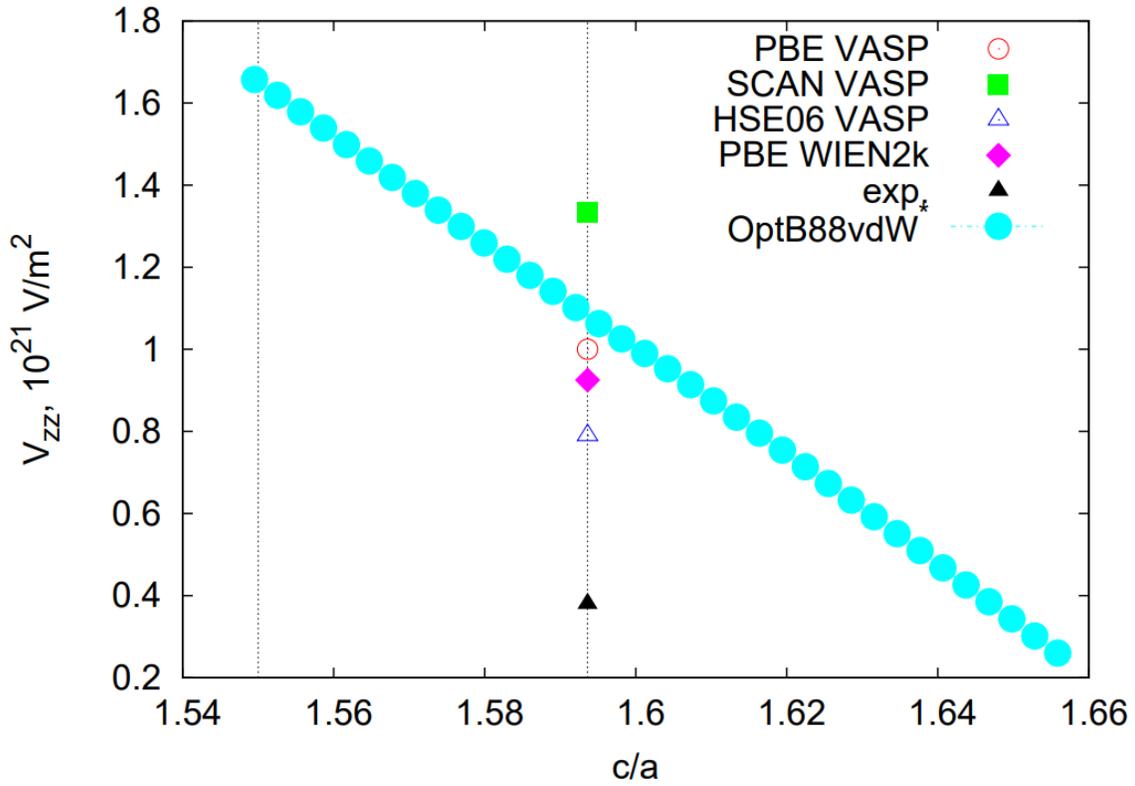

*Fig. 1 This plot illustrates the sensitivity of the $V_{zz}$ in the situation when the nearest neighbor environment in nearly cubic, and $V_{zz}$ is nonzero but numerically small for Scandium case (JVASP-996). The leftmost dotted vertical line corresponds to the JARVIS-DFT-OptB88vdW calculated c/a and the rightmost vertical line corresponds to the experimentally determined c/a. It also illustrates the effect of choosing different energy functionals and band structure methods. PBE is the standard gradient-corrected density functional (GGA), OptB88vdW denote the calculations*

*performed as describe in the paper for the entire database (in the rest of the paper, no attempt was made to correct the c/a ratio), SCAN is a recently proposed meta-GGA functional, which improves the results for strongly correlated systems (which Sc is not), HSE06 is a hybrid nonlocal functional (see VASP manual for details) and WIEN2k is an all-electron LAPW code, also utilizing PBE. Gaussian integration was used throughout the paper, except this figure, where we used tetrahedrons to reduce noise and elucidate a clean linear dependence.*

The second caveat is that, due to the nature of our massive calculations, we could not manually introduce experimental magnetic order in each case (and, in many cases, it is simply not known). To this end, we did blanket calculations for each material starting from a ferromagnetic (FM) configuration of the electron spin. In most, albeit not all cases, such procedure converges to an FM solution even for experimentally antiferromagnetic (AF) materials and captures the effect of the local magnetization on the EFGs at least semi-quantitatively. Notable exceptions occur when AF order triggers specific orbital ordering, such as Kugel-Khomskii effect[49]. Another dramatic example is provided by Fe-based superconductors, where the observed magnetic ordering breaks the tetragonal symmetry. Calculations show[50] that the FM and the Neel AF state (which does not break the tetragonal symmetry) have a relatively minor (10% - 30%) effect on the EFGs. For these same materials, but with proper stripe-like (symmetry-breaking) magnetic ordering imposed, the EFG changes by up to an order of magnitude. This caveat should be kept in mind when using the database.

Finally, for heavy elements, like uranium, one may expect that including spin-orbit coupling may alter the results. However, so far our experience has been that even in those cases the effect of spin-orbit[51] is at best moderate, so this does not appear to be a serious limitation.

With this in mind, let us proceed to an experimental validation for selected cases where experimental benchmarks were available. As there is no systematic chemistry and spacegroup based experimental database for electric field gradients, we manually search for experimental data and also extract values from the NQRS database from JAICI[37,38]. Since neither NQRS, nor the original references, necessarily give information on the crystal symmetry, we ensure that the proper polymorphs are being compared in cases of uncertainty by choosing materials that have only one known polymorph. In addition, we verify that the number of $v_Q$'s matched the number of expected inequivalent sites. Furthermore, we only extract experimental values determined at room temperature or colder; in most cases below 100 K.

After extracting the experimental values, we compared them to our DFT predictions. The comparisons are shown in Table 2. Details of each material are available in the database with its corresponding webpage, for example, https://www.ctcms.nist.gov/~knc6/jsmol/JVASP-4149 for JARVIS-ID: JVASP-4149. The list of materials for experimental comparison consists of several chemical classes such as 17 unary, 16 binary, 5 ternary compounds including oxides, sulfides, nitrides, halides, and gallides. In addition to the experimental data, we compare with previously computed data. The mean absolute deviation of the experimental comparison dataset is $1.17 \times 10^{21} Vm^{-2}$, while the mean absolute percentage difference is 28.91 %. The absolute deviation varies from a low value of $0.03 \times 10^{21} Vm^{-2}$ for titanium to a maximum value of $5.3 \times 10^{21} Vm^{-2}$ for $UAs_2$.

Our DFT and experimental data show close agreements (Pearson's coefficient 0.999) as shown in Fig. 2a. Our computational data compares well with the previously reported values (Pearson's coefficient 0.999) as also shown in Fig. 2b. The previously reported data could be either from FLAPW or PAW based calculations and different pseudopotentials suggesting that different computational methods have only a moderate impact on the predicted values. As discussed, materials close to a local cubic environment, such as Sc, exhibit high sensitivity to the crystal structure, in those cases, to *c/a*. Using the experimental *c/a*, rather than the calculated value, greatly improves the agreement with the experiment; see Fig. 1. Even without that, however, the overall computational predictions compare well with experiments, as seen in Fig. 2.

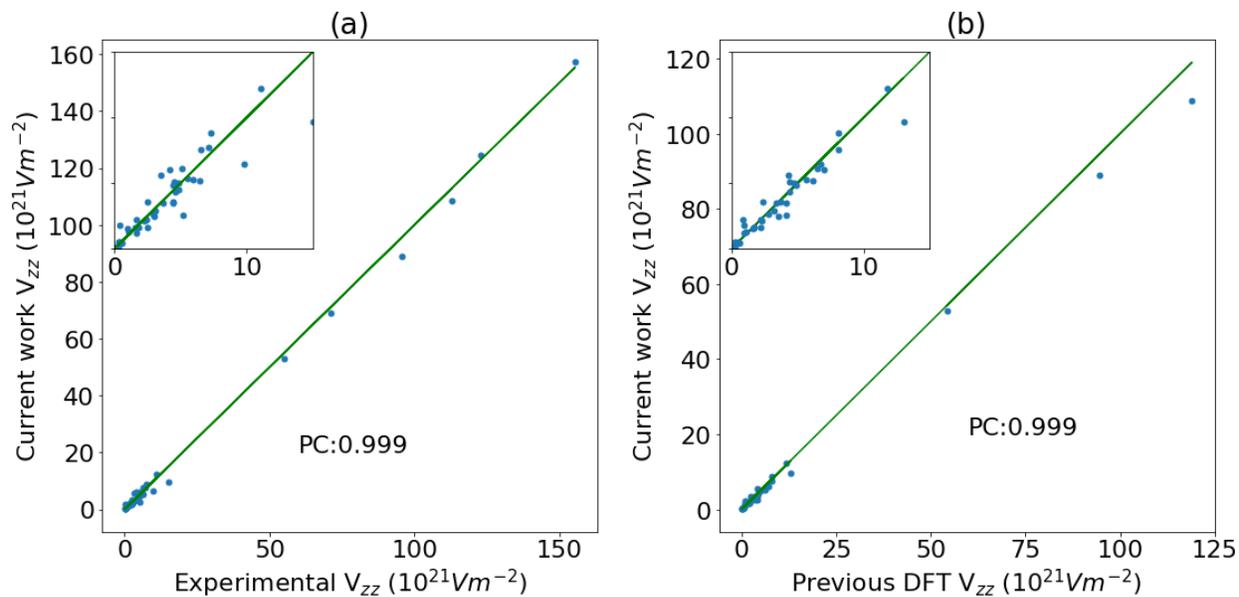

*Fig. 2 Comparison of current-work, experimental, and previous calculation data from Table. 2. a) experimental data vs DFT data from current work, b) previously computed data vs current work data. The lines are the ideal case with a slope of unity.*

*Table 2 Comparison of current density functional (J-DFT) predictions with experimental (Exp) and previously (Prev.-DFT) reported Electric Field Gradient, $V_{zz}$ ($10^{21}Vm^{-2}$) data. The MAD (Mean Absolute Deviation), and MAPD (Mean Absolute Percentage Difference) values are calculated for the whole data. Details of each material are provided at its corresponding webpage. Please look into the references (and references therein) for experimental and previously calculated data.*

| Material | JID | Atom | $\|V_{zz\,(Exp)}\|$ | $\|V_{zz\,(J\text{-}DFT)}\|$ | $\|V_{zz\,(Prev.\text{-}DFT)}\|$ | $\|\Delta\|$ | $\|\Delta\%\|$ |
|---|---|---|---|---|---|---|---|
| $Cl_2$ | 855 | Cl | 55.18[52] | 52.85 | 54.23[41] | 2.33 | 4.22 |
| $Br_2$ | 840 | Br | 95.69[52] | 88.86 | 94.44[41] | 6.83 | 7.14 |
| $I_2$ | 895 | I | 113.00[52] | 108.70 | 119.01[41] | 4.30 | 3.81 |
| Be | 25056 | Be | 0.044[53] | 0.072 | 0.06[54] | 0.028 | 63.64 |
| Mg | 14840 | Mg | 0.048[53] | 0.079 | 0.04[54] | 0.031 | 64.58 |
| Sc | 996 | Sc | 0.38[53] | 1.78 | 0.96[54] | 1.40 | 368.4 |
| Ti | 14815 | Ti | 1.61[53,55] | 1.64 | 1.75[54] | 0.03 | 1.86 |
| Co | 858 | Co | 2.9[53] | 0.52 | 0.29[54] | 2.38 | 82.06 |
| Zn | 1056 | Zn | 3.48[53] | 5.62 | 4.29[54] | 2.14 | 61.50 |
| Zr | 14612 | Zr | 4.40[53] | 3.50 | 4.14[54] | 0.90 | 20.45 |
| Tc | 1020 | Tc | 1.83[53] | 1.67 | 1.74[54] | 0.16 | 8.74 |
| Ru | 987 | Ru | 0.97[53] | 1.52 | 1.62[54] | 0.55 | 56.70 |
| Cd | 14832 | Cd | 6.50[53] | 7.56 | 8.13[54] | 1.06 | 16.31 |
| La | 910 | La | 1.62[53] | 2.24 | 0.91[54] | 0.62 | 38.27 |
| Hf | 14590 | Hf | 7.33[53] | 8.87 | 8.12[54] | 1.54 | 21.01 |
| Re | 981 | Re | 5.12[53] | 6.14 | 6.49[54] | 1.02 | 19.92 |
| Os | 952 | Os | 4.16[53] | 6.00 | 7.02[54] | 1.84 | 44.23 |
| $BI_3$ | 3630 | I | 71.29[56] | 68.98 | - | 2.31 | 3.24 |
| $CF_3I$ | 32512 | I | 124.34[57] | 123.22 | - | 1.12 | 0.90 |
| ClN | 5758 | I | 157.21[58] | 151.0 | - | 6.21 | 3.95 |
| $NaNO_2$ | 1429 | Na | 0.438[59] | 0.552 | 0.575[60] | 0.114 | 26.03 |
| $NaNO_2$ | 1429 | N | 11.10[59] | 12.194 | 11.772[60] | 1.094 | 9.86 |
| $Cu_2O$ | 1216 | Cu | 9.80[61] | 6.47 | 6.765[60] | 3.33 | 33.98 |
| $TiO_2$ | 10036 | Ti | 2.21[62] | 2.098 | 2.269[60] | 0.112 | 5.07 |
| $TiO_2$ | 10036 | O | 2.38[62] | 2.21 | 2.235[60] | 0.17 | 7.14 |
| $SrTiO_3$ | 8082 | O | 1.62[63] | 1.24 | 1.00[64] | 0.38 | 23.46 |
| $BaTiO_3$ | 8029 | O | 2.46[63] | 3.56 | 2.35[64] | 1.10 | 44.72 |
| $Li_3N$ | 1375 | N | 1.04[65] | 1.25 | 1.09[60] | 0.21 | 20.19 |
| $Li_3N$ | 1375 | Li(2c) | 0.30[65] | 0.225 | 0.291[60] | 0.075 | 25.00 |
| $Li_3N$ | 1375 | Li(1b) | 0.60[65] | 0.50 | 0.616[60] | 0.144 | 24.00 |
| FeSi | 8178 | Fe | 4.45[66,67] | 4.84 | 4.92[41] | 0.39 | 8.76 |
| $FeS_2$ (marcasite) | 2142 | Fe | 3.0[68] | 2.93 | 3.21[41] | 0.07 | 2.33 |
| $FeS_2$ (pyrite) | 9117 | Fe | 3.66[68] | 3.51 | 3.40[41] | 0.15 | 4.10 |
| $2H-MoS_2$ | 54 | Mo | 7.09[37] | 7.70 | - | 0.61 | 8.60 |
| $2H-MoS_2$ | 54 | S | 5.54[37] | 5.33 | - | 0.21 | 3.80 |
| 2H-WS$_2$ | 72 | S | 4.82[37] | 4.53 | - | 0.29 | 6.22 |
| CaGa$_2$ | 16464 | Ga | 4.44[69] | 3.55 | 3.77[70] | 0.89 | 20.05 |
| SrGa$_2$ | 14853 | Ga | 5.22[69] | 2.54 | 4.13[70] | 2.68 | 51.34 |
| BaGa$_2$ | 19628 | Ga | 4.48[69] | 5.10 | 4.38[70] | 0.62 | 13.84 |
| NaGa$_4$ | 14728 | Ga(e) | 6.49[70] | 5.20 | 6.18[70] | 1.29 | 19.88 |
| NaGa$_4$ | 14728 | Ga(d) | 4.64[70] | 4.33 | 4.44[70] | 0.31 | 6.68 |
| CaGa$_4$ | 20533 | Ga(e) | 2.89[70] | 2.67 | 2.80[70] | 0.22 | 7.61 |
| CaGa$_4$ | 20533 | Ga(d) | 4.87[70] | 4.99 | 4.73[70] | 0.12 | 2.46 |
| SrGa$_4$ | 20206 | Ga(e) | 2.51[70] | 1.67 | 2.24[70] | 0.84 | 33.47 |
| SrGa$_4$ | 20206 | Ga(d) | 5.95[70] | 5.31 | 5.64[70] | 0.64 | 10.76 |
| TaP | 79643 | Ta | 3.00[71] | 2.5 | 3.54[71] | 0.50 | 16.67 |
| UAs$_2$ | 19797 | U | 15.0[72] | 9.7 | 13.03[51] | 5.3 | 35.3 |
| **MAD** | | | | | | 1.17 | - |
| **MAPD** | | | | | | - | 28.90 |

Next, we present some statistical analysis, which utilizes the unprecedently large scope of this study. In Fig. 3 we present the histograms for the principal components of EFG for 15187 materials. As shown in Fig. 3, it is interesting to observe that our initial weeding out of the locally cubic sites was incomplete, because ~30% of sites have a zero EFG. It is partially due to the fact that in many compounds some, but not all sites are locally cubic.

The EFG component distributions exponentially decay as we approach high values suggesting that there are relatively few materials with high EFG. Similar behavior is observed for all three EFG components. The $V_{zz}$ distribution range (Fig. 3c) is higher than the other two components because by convention $V_{zz}$ is the highest EFG value. We find that some of the high-$V_{zz}$ material examples are: I in S(IO$_3$)$_2$ (JVASP-5821):185.93; IBr (JVASP-2029):162.7, Bi in K$_2$BiRb (JVASP-80254): 130.3, W in Na$_4$O$_4$W (JVASP-42416):95.7, Pt in NaH$_4$Pt (JVASP-22691): 92.9, and Sb in BaBr$_2$Sb (JVASP-65626):74.80. Interestingly, all of these very high-$V_{zz}$ materials are vdW-bonded which could be responsible for high electron cloud asymmetry. The asymmetry parameter, $\eta$ is calculated using Eq. 1 and we observe that majority of the atomic sites have zero asymmetry parameters by symmetry, as shown in Fig. 3d.

In order to find the highest $\eta$ sites, we screen for materials with $\eta \approx 1$ (note that no site symmetry can lead to $\eta$ being exactly 1, so such materials simply happen to have one of the principal components of the EFG tensor numerically small), which leads to candidates such as Te in Rb$_2$Te$_5$ (JVASP-4149), and O in ZnTiO$_3$ (JVASP-11167). Some more examples are shown in the Table. 3. The list consists of compounds with diverse chemistry. In addition to the EFG and materials information, the table contains important information, such as OptB88vdW bandgaps and energy above the convex hull that might be important from the experimental perspectives. The complete dataset of the EFG tensor and the asymmetry parameter is provided in the data-records section, from which several similar candidate materials can be easily identified.

*Table. 3 Some example compounds with high asymmetry parameter ($\eta \approx 1$). Chemical formula, JARVIS-ID, bandgap ($E_g$)(eV), energy above convex hull ($E_{hull}$) (eV), electric field gradients ($V_{ZZ}$, $V_{YY}$) ($10^{21}Vm^{-2}$) and the asymmetry parameter ($\eta$) information is provided.*

| **Materials** | **Atom** | **JID** | **$E_g$ (eV)** | **$E_{hull}$ (eV)** | **$V_{ZZ}$** | **$V_{YY}$** |
|---|---|---|---|---|---|---|
| **Rb$_2$Te$_5$** | Te | 4149 | 0.22 | 0.0 | 58.07 | -58.06 |
| **InPS$_4$** | S | 3465 | 2.33 | 0.0 | 16.86 | -16.83 |
| **NbTe₂** | Nb | 7720 | 0.00 | 0.0 | -4.08 | 4.07 |
| **ZnTiO₃** | O | 11167 | 3.05 | 0.0 | 2.97 | -2.97 |
| **Na₂Hf₂O₅** | O | 41602 | 2.15 | 0.0 | 3.14 | -3.14 |
| **CaPt₅** | Pt | 18644 | 0.0 | 0.0 | 21.12 | -21.05 |
| **PbCO₃** | O | 32164 | 3.42 | 0.0 | -11.08 | 11.05 |
| **BiNaSe₂** | Se | 8817 | 0.76 | 0.0 | 22.13 | -22.00 |
| **KNbO₃** | Nb | 8083 | 1.88 | 0.002 | 2.22 | -2.20 |
| **AlPO₄** | P | 13743 | 5.44 | 0.04 | 1.79 | -1.79 |
| **NbBiO₄** | Nb | 40901 | 2.47 | 0.09 | -6.43 | 6.42 |

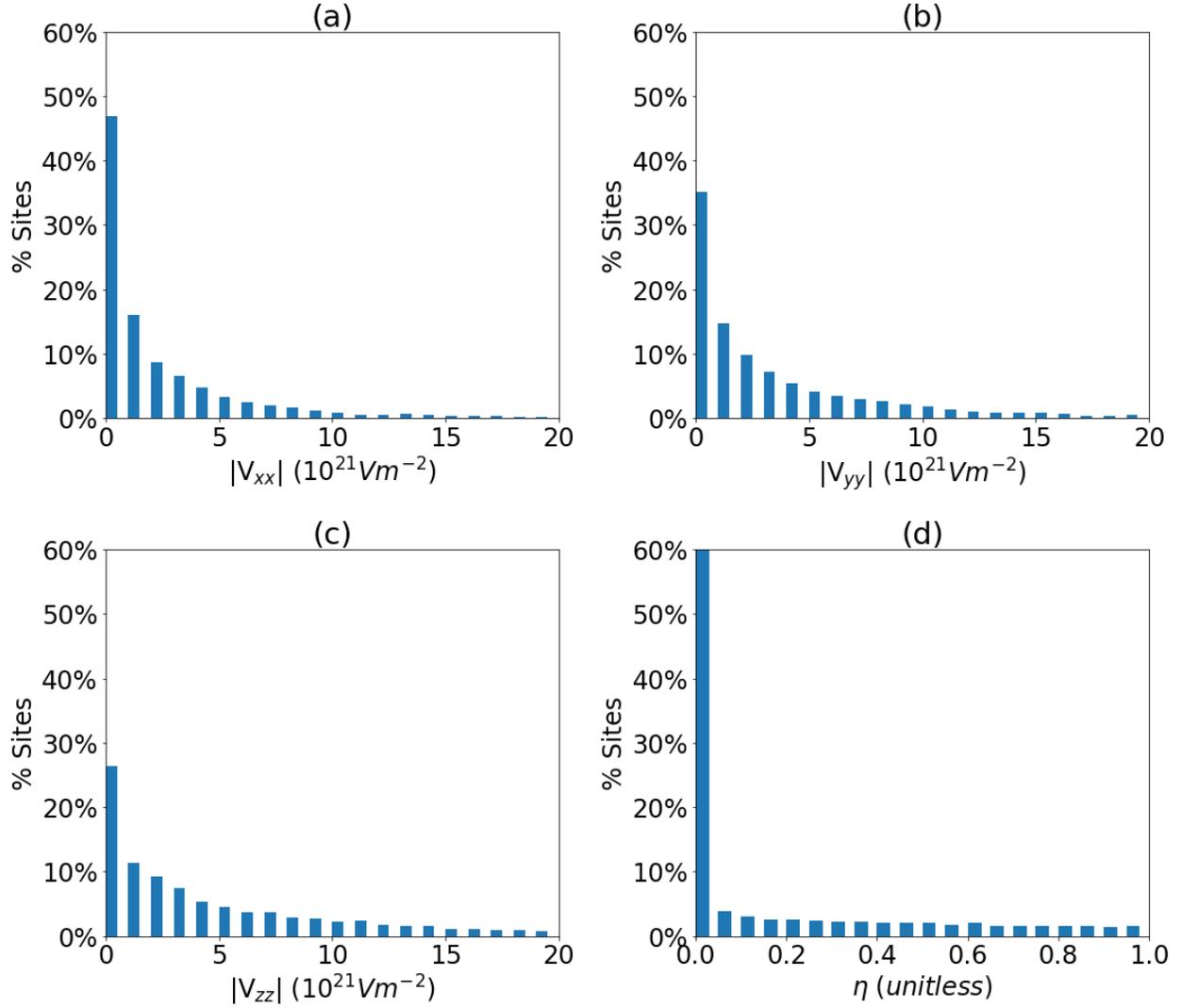

*Fig. 3 EFG tensor component and asymmetry parameter distribution for all the atomic sites in the materials under investigation. a) $V_{XX}$, b) $V_{YY}$, c) $V_{ZZ}$, d) asymmetry parameter. In the panel (d), sites in which $V_{zz} = 0$, and therefore $\eta$ is ill-defined, are excluded from the dataset.*

Next, we identify correlations in the EFG parameters in Figs. 4 and 5. In Fig. 4a we plot the numerator and denominator of the asymmetry parameter equation. Clearly in Fig. 4a, if the numerator is zero, the asymmetry parameter is zero, while atomic sites lying on the $x = y$ line represents the highest asymmetric parameter ($\eta = 1$) sites; the later may be of interest from an experimental perspective. Obviously, the highest $\eta$ is obtained when $V_{xx}$ is zero but $V_{zz}$ is not. In Fig. 4b, we plot the asymmetry parameter against the $V_{zz}$ component which shows a bell-shape feature. Upon initial appearance, Fig. 4b suggests that high asymmetry behavior is preferentially observed in low $V_{zz}$ values. However, the 3D histogram (shown in Fig. 5), which is plotted on a logarithmic scale, reveals a similar distribution of $\eta$ values for all $V_{zz}$; there are simply much fewer materials with high $V_{zz}$.

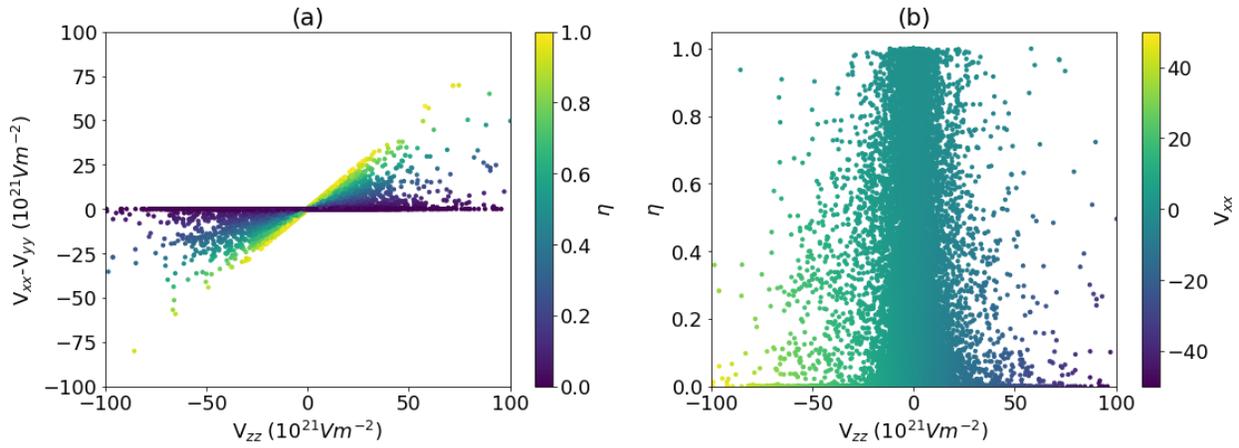

*Fig. 4 Relation between EFG components and the asymmetry parameter. a) $V_{xx}$ -$V_{yy}$ vs. $V_{zz}$, b) $V_{zz}$ vs the asymmetry parameter vs. the $V_{zz}$.*

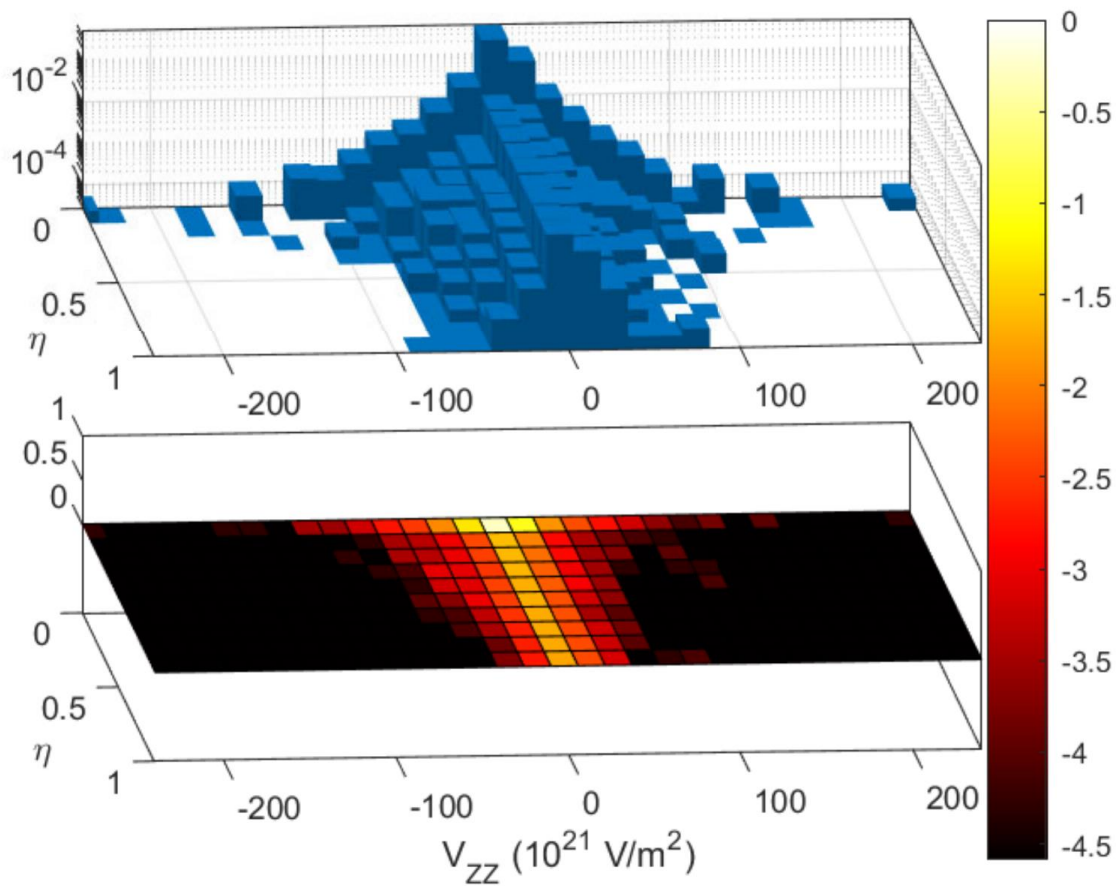

*Fig. 5 Histogram (above) and intensity map (below) of the probability are shown as a function of $V_{ZZ}$ and $\eta$. The logarithmic z-scale allows for the strong preponderance of small $V_{ZZ}$ and small $\eta$ values to be clearly seen, while at the same time revealing the bell-shaped distribution in $V_{ZZ}$ for all $\eta$ values. Values where $V_{zz}=0$, and therefore $\eta$ is ill-defined, have been excluded.*

In Fig. 6, we show the periodic table trends for the elements with their highest $V_{zz}$ value among all the materials under investigation, containing this element. Interestingly, we observe that halides such as Br and I, transition elements such as Au and Pt, actinides such as U, pnictides such as Bi and Sb, and inert gases such as Xe attain high $V_{zz}$ values. Low atomic weight elements such as H, Be and Li have low $V_{zz}$ values. These results can be important for experiments because it represents the overall trends during NMR/NQR frequencies. We note that these trends are not for individual elemental systems but elemental distribution in the multicomponent systems.

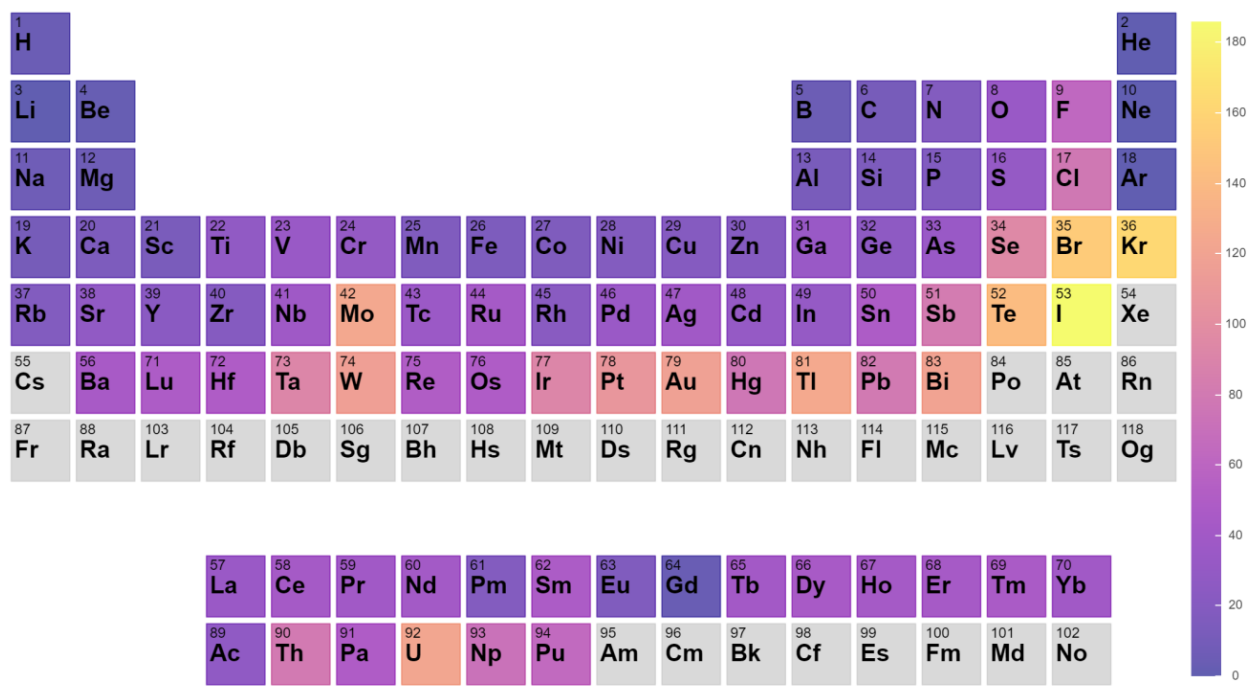

*Fig. 6 Periodic table trends for elements with their highest $V_{zz}$ in all possible systems in the current work.*

**Usage notes**

The database presented here represents the largest collection of consistently calculated electric field gradient properties of materials using density functional theory assembled to date. We anticipate that this dataset, and the methods provided to access it, will provide a useful tool in fundamental and application-related studies of materials. Our actual experimental verification provides insight into understanding the applicability and limitation of our DFT data. Based on the list of data, the user will be able to choose particular materials for specific applications. Data mining, data analytics, and artificial-intelligence tools then can be added to guide the design and optimization of materials.

**Code availability**

Python-language based codes for carrying out calculations and analyzing the results are provided at the github page: https://github.com/usnistgov/jarvis .

**Contributions**

All authors jointly developed the workflow. KC carried out the high-throughput DFT calculations. JNA, IM, KLS helped in the experimental validation section. All contributed in writing the manuscript.

**Acknowledgements**

K.C. thanks the National Institute of Standards and Technology for funding, computational and data-management resources. K.C. also thank the computational support from XSEDE computational resources under allocation number (TG-DMR 190095). JNA acknowledges support from the Quantum Science and Engineering Center at George Mason University as well as the National Science Foundation (ECCS-1711118).

**Competing interests**

The authors declare no competing interests.